\documentclass[%
 reprint,
superscriptaddress,
 amsmath,amssymb,
 aps,
prl
]{revtex4-2}

\usepackage{graphicx}
\usepackage{dcolumn}
\usepackage{bm}

\usepackage{xcolor}

\begin{document}

\preprint{APS/123-QED}

\title{Subwavelength confinement of optical and acoustic waves for enhancing Brillouin gain in a planar waveguide structure}

\author{A. R. Mukhamedyanov}
\affiliation{Moscow Institute of Physics and Technology, 141700, 9 Institutskiy pereulok, Moscow, Russia}
\affiliation{Dukhov Research Institute of Automatics (VNIIA), 127055, 22 Sushchevskaya, Moscow, Russia}

\author{E. A. Tereshchenkov}
\affiliation{Moscow Institute of Physics and Technology, 141700, 9 Institutskiy pereulok, Moscow, Russia}
\affiliation{Dukhov Research Institute of Automatics (VNIIA), 127055, 22 Sushchevskaya, Moscow, Russia}
\affiliation{Institute for Theoretical and Applied Electromagnetics, 125412, 13 Izhorskaya, Moscow, Russia}

\author{E. S. Andrianov}
\affiliation{Moscow Institute of Physics and Technology, 141700, 9 Institutskiy pereulok, Moscow, Russia}
\affiliation{Dukhov Research Institute of Automatics (VNIIA), 127055, 22 Sushchevskaya, Moscow, Russia}
\affiliation{Institute for Theoretical and Applied Electromagnetics, 125412, 13 Izhorskaya, Moscow, Russia}

\author{A. A. Zyablovsky}
\email{zyablovskiy@mail.ru}
\affiliation{Moscow Institute of Physics and Technology, 141700, 9 Institutskiy pereulok, Moscow, Russia}
\affiliation{Dukhov Research Institute of Automatics (VNIIA), 127055, 22 Sushchevskaya, Moscow, Russia}
\affiliation{Institute for Theoretical and Applied Electromagnetics, 125412, 13 Izhorskaya, Moscow, Russia}

\date{\today}

\begin{abstract}

We propose a planar waveguide structure consisting of two layers separated by a gap that provides subwavelength confinement of optical and acoustic modes and a large overlap integral between them. This, in turn, leads to an increase in Brillouin gain. Using the structure of two Si$_3$N$_4$ layers with a subwavelength gap from SiO$_2$ as an example, we demonstrate that a hybridized transverse-magnetic (TM) optical mode exhibits strong field confinement in the SiO$_2$ gap. Simultaneously, a high-frequency acoustic mode can be confined in the same gap, providing a high spatial overlap between the optical and acoustic fields. This provides a high opto-acoustic interaction with a Brillouin gain up to $5$ W$^{-1}$m$^{-1}$. The proposed structure is compatible with planar photonic integration and offers a promising platform for compact, low-threshold Brillouin lasers and coherent phonon generation.

\end{abstract}

\maketitle

\section{Introduction}
Brillouin scattering is the effect when an acoustic wave scatters an electromagnetic waves. In integrated photonics Brillouin scattering has a wide range of applications, from ultra-narrow linewidth lasers \cite{li2012characterization, panyaev2024sub, ye2025integrated} and microwave-optical frequency conversion \cite{chen2023optomechanical, yang2024proposal} to sensing \cite{godet2022micronewton, huang2017brillouin, zhang2022refractive, mukhamedyanov2023subthreshold} and non-reciprocal light propagation~\cite{dong2015brillouin, sohn2021electrically}. The Brillouin laser finds application in integrated photonics as a nonlinear element \cite{eggleton2019brillouin, ye2025integrated, slinkov2025all, mukhamedyanov2026using, mukhamedyanov2026all}.  The magnitude of Brillouin interaction is quantified by Brillouin gain $G_B$, which is determined by the acoustic properties of the material and by the distribution and mutual overlap of optical and acoustic waves. The gain determines the intensity of the Stokes wave and, among other things, the threshold of Brillouin lasers based on that structure \cite{wiederhecker2019brillouin, wolff2015, mukhamedyanov2024hard, mukhamedyanov2025control}. This motivates substantial effort to its increasing in a wide variety of optical structures in past decades \cite{eggleton2019brillouin, merklein2022100, ren2025stimulated}.

The gain associated with photoelastic forces is inversely proportional to the volume of the acoustic mode and proportional to the overlap integral of the optical and acoustic modes \cite{qiu2013stimulated, wiederhecker2019brillouin}. Therefore, to increase gain, it is necessary to achieve strong confinement of the optical and acoustic modes in the material with a large photoelastic density force. However, confining both optical and acoustic modes in the same region is a nontrivial task, because the same combination of materials must meet both optical and acoustic confinement requirements. For example, SiO$_2$ is a photoelastic material, but in integrated photonics SiO$_2$ serves as a cladding around a higher-index core, e.g., Si or Si$_3$N$_4$, with a lower photoelastic coefficient. To increase Brillouin gain for such geometry, the optical modes must extend into the SiO$_2$ substrate. For example, in the case of one strip ${\rm{Si_3 N_4}}$ waveguide embedded in ${\rm{SiO_2}}$ \cite{gundavarapu2019, liu2025large, chauhan2021visible}, it is necessary to use sub $50~{\rm{nm}}$ waveguide thick with over $\sim 10~{\rm{\mu m}}$ cladding width. However, fabrication of such thin and wide waveguides is a challenging problem \cite{jin2021hertz}. In thin-core Si$_3$N$_4$ waveguides, weak optical confinement can require large bend radii to keep losses low.  This limits the scalability and integration of the nitride-based devices. Furthermore, for such a design, acoustic energy can leak into the surrounding SiO$_2$ cladding, limiting the peak Brillouin gain.

Recently, several approaches to enhancing Brillouin gain have been proposed \cite{eggleton2019brillouin, merklein2022100, ren2025stimulated, rakich2012giant, liu2025large}. Such approaches are based on the confinement of the acoustic modes in spatially confined structures, such as suspended waveguides \cite{yu2025chip, kittlaus2016large, GonzalezAndrade2024,NunoRuano2025, VanLaer2015}. Using such structures makes it possible to limit the leakage of acoustic modes into the substrate, achieve their spatial confinement, and increase the magnitude of the opto‑acoustic interaction. The advantage of this platform is the removal of the silica under-cladding that enables the suppression of the phonon leakage, which limits Brillouin interactions. However, despite its advantages, these structures have significant fabrication complexity \cite{cao2018surface} and require isolation of the device from heat-spreading and electronic layers. Properties of such systems are highly sensitive to sidewall roughness and fabrication processes.

Another approach is based on using photonic and optomechanical crystals \cite{zurita2021,Dainese2006, mayor2025high, eichenfield2009optomechanical, jiang2019lithium}, which are used for guidance of both optical and acoustic modes. The use of such structures is limited by the complexity of their fabrication, and in the case of two‑dimensional crystals, by the difficulty of using them in integrated photonics with planar waveguides. Therefore, planar waveguide structures~\cite{gundavarapu2019,botter2022guided} are more convenient in terms of integration on a chip and fabrication simplification.

In this work, we propose a planar optical waveguide structure that provides subwavelength localization of both optical and acoustic modes to enhance the Brillouin gain. Our structure consists of two layers separated by a nanoscale gap. The refractive indices and acoustic velocities of the materials are selected in such a way as to support waveguide optical and acoustic modes. Using the structure of two  Si$_3$N$_4$ layers with a gap from SiO$_2$ as an example, we demonstrate that there is a waveguide geometry where optical and acoustic modes are confined to a subwavelength area. The subwavelength confinement, combined with a high mode overlap factor, leads to a significant increase in the Brillouin gain, which is up to $5$ W$^{-1}$m$^{-1}$. The obtained results offer a promising pathway for developing compact, low-power Brillouin-based devices for integrated photonics.

\section{Approach for subwavelength confinement of optical and acoustic modes in a waveguide planar structure}

Increasing the interaction between optical and acoustic fields is a critical practical task. The quantity that characterizes the strength of such interaction is the Brillouin gain:
\begin{equation}
 G_B(\Omega)=\frac{\omega_s}{\Omega}
 \frac{P_{\mathrm{ac}}(\Omega)}{P_pP_s},
 \label{eq:1}
\end{equation}
where $P_p$ and $P_s$ are the time-averaged powers in the pump's and Stoke's optical modes. $\Omega$ and $\omega_s$ are the frequencies of the acoustic and the Stokes modes, respectively. The mechanical power delivered per unit length is \cite{qiu2013stimulated,wolff2015}
\begin{equation}
 \begin{aligned}
 P_{\mathrm{ac}}(\Omega)&=\frac{\Omega}{2}\operatorname{Im}\bigg[
 \iint\mathbf f_{\mathrm{pe}}^*\cdot\mathbf u(\Omega)\,dA\\
 &\qquad+\oint\mathbf f_{\mathrm{mb}}^*\cdot\mathbf u(\Omega)\,dl\bigg],
 \end{aligned}
 \label{eq:2}
\end{equation}
Here $\mathbf u$ is the acoustic displacement field; $f_{\mathrm{pe, i}} = -\frac{1}{2} \varepsilon_0 n_{\text{eff}}^4 \partial_{j}~p_{ijkl} E_k E_l$ is the photoelastic density force \cite{wiederhecker2019brillouin, van2016unifying} with $p_{ijkl}$ - the fourth-rank photoelastic tensor. $\mathbf f_{\mathrm{mb}} = \hat{\mathbf{n}} (\varepsilon_0\delta\varepsilon_{mb} (\textbf{E}_{p, \parallel}^* \cdot \textbf{E}_{s, \parallel}) + \varepsilon_0^{-1} \delta\varepsilon_{mb}^{-1} (\textbf{D}_{p, \perp}^* \cdot \textbf{D}_{s, \perp}))$ is the moving-boundary contribution \cite{rodrigues2023stimulated, rodrigues2025cross}. Here $\hat{\mathbf{n}}$ is the unit normal vector. The vector $\mathbf{D}_{i,\perp}$ is the normal electric displacement field and the vector $\mathbf{E}_{i,\parallel}$ is the tangential electric field. The $\delta \varepsilon_{mb} = \varepsilon_{1} - \varepsilon_{2}$ is the jump in the dielectric permittivity at the boundary between different materials.

From the expression for $f_{pe}$ and $G_B$ can be obtained that the gain is proportional to the overlap integral between the optical and acoustic modes and inversely proportional to the volume of the acoustic mode \cite{qiu2013stimulated}. Consequently, enhancement of the opto‑acoustic interaction can be achieved in the structures where there is a subwavelength confinement of the acoustic and optical modes in the same volume.

\begin{figure}[htbp]
\centering\includegraphics[width=0.8\linewidth]{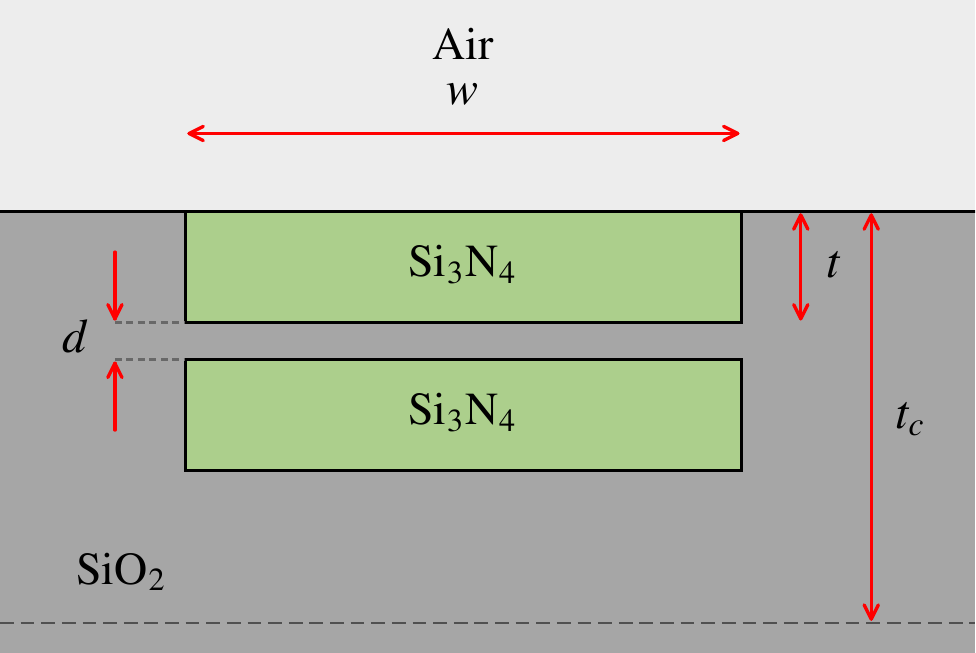}
\caption{Scheme of the waveguide structure.}. 
\label{fig:1}
\end{figure}

For this purpose, we consider a waveguide consisting of two identical layers separated by a nanoscale gap, as shown in Fig.~\ref{fig:1}. The refractive index of the layers $n_{1}$ is greater than the refractive index of the gap $n_{2}$. In such a system there can be hybrid TM optical modes with effective index $n_{2} < n_{\mathrm{eff}} < n_{1}$. When this inequality is satisfied, the electric field amplitude near the boundary of the two media is amplified by an amount $E_{\perp,1}/E_{\perp,2} = n_{1}^2/n_{2}^2 >1$. In this case, the electric field in the gap is the sum of two evanescent waves, which have maxima at the boundaries between the gap and the layer. This leads to the overlap of the two evanescent fields across the gap. The overlap of two TM optical modes produces optical gap confinement of the hybrid TM mode~\cite{almeida2004guiding}.

For acoustic waves the confinement requirements differ from those of optical ones. Longitudinal acoustic waves are guided when $v_{L,2}< v_m < v_{L,1}$ \cite{poulton2013acoustic, wolff2015}, where $v_m$ is the acoustic mode velocity, $v_{L,2}$ and $v_{L,1}$ are the bulk longitudinal acoustic velocities in the gap and the layer, respectively. 

If the specified conditions for the refractive indices and acoustic velocities of the materials are simultaneously met, it is possible to achieve the subwavelength confinement of optical and acoustic modes in the gap. This confinement, both acoustic and optical modes in the gap, can achieve high Brillouin gain. To confirm this concept, we investigate a waveguide structure based on SiO$_2$ and Si$_3$N$_4$ that demonstrates such a confinement and a high gain.

\section{Example of the planar waveguide structure based on two Si$_3$N$_4$ layers}

We consider a system consisting of two Si$_3$N$_4$ layers in a SiO$_2$ substrate, which are placed at a distance $d$, creating in SiO$_2$ a gap region. The schematic of this system is shown in Fig.~\ref{fig:1}. The high refractive index differences between Si$_3$N$_4$ and the surrounding material lead to a confinement optical mode in the gap of the substrate SiO$_2$ ($n_{\text{SiO}_2} < n_{\mathrm{eff}} < n_{\text{Si}_3\text{N}_4}$). Furthermore, due to the difference in the material density, the longitudinal acoustic velocities satisfy the inequality $v_{L,\text{SiO$_2$}} = 5.97 \text{ km/s} < v_m < 10.23 \text{ km/s} = v_{L,\text{Si$_3$N$_4$}}$. This means that the acoustic mode is confined in the SiO$_2$. The photoelastic coefficients are higher in SiO$_2$ than in Si$_3$N$_4$ therefore interaction between acoustic and optical modes is higher Table~\ref{tab:2}.

We calculate the Brillouin gain from the optically driven acoustic response~\cite{wolff2015}. For a pump and a counterpropagating Stokes wave, $\Omega=\omega_p-\omega_s$ and $\beta_m=\beta_p-\beta_s\simeq4\pi n_{\mathrm{eff}}/\lambda_p$ \cite{merklein2022100}, where $\omega_{p,s}$ and $\beta_{p,s}$ are the frequencies and the wavenumbers of the pump and Stokes waves, $\lambda_p$ is the pump wavelength, $\beta_s<0$. The full vector acoustic equation is solved at real $\Omega$ and fixed $\beta_m$ using the finite-element method with perfectly matched layers. The photoelastic contribution is calculated in weak stress form. The moving-boundary contribution is calculated on the waveguide boundaries. The material properties used in our finite-element simulations are summarized in Table~\ref{tab:2}. For the material SiO$_2$, we use the stiffness matrix elements $C_{ij}$ and the appropriate photoelastic coefficients Table~\ref{tab:2}.

\begin{table*}[htbp]
\centering
\begin{tabular}{|c|c|c|c|c|c|}
\hline
        Material & Refractive Index, $n$ & Density, $\rho$ kg/m$^3$ & Young's modulus, GPa & Poisson's ratio & Photoelastic coefficients \\ \hline
        Si$_3$N$_4$ & 2 & 3100 & 280 & 0.23 & $p_{11}=-0.125$, $p_{12}=0.047$ \\ \hline
        SiO$_2$ & 1.45 & 2203 &  \multicolumn{2}{|c|}{$\{C_{11} = 78.5, ~C_{12} = 16.1\}$} & $p_{11}=0.121$, $p_{12}=0.271$ \\ \hline
\end{tabular}
\caption{Material parameters and photoelastic coefficients used in the simulations. For SiO$_2$ and Si$_3$N$_4$, the elastic constants are given in Voigt notation. Photoelastic coefficients are taken from \cite{gyger2020observation,mishra2026ultra}. The isotropic relation $p_{44}=(p_{11}-p_{12})/2$ is used.}
\label{tab:2}
\end{table*}

\begin{figure}[htbp]
\centering\includegraphics[width=1\linewidth]{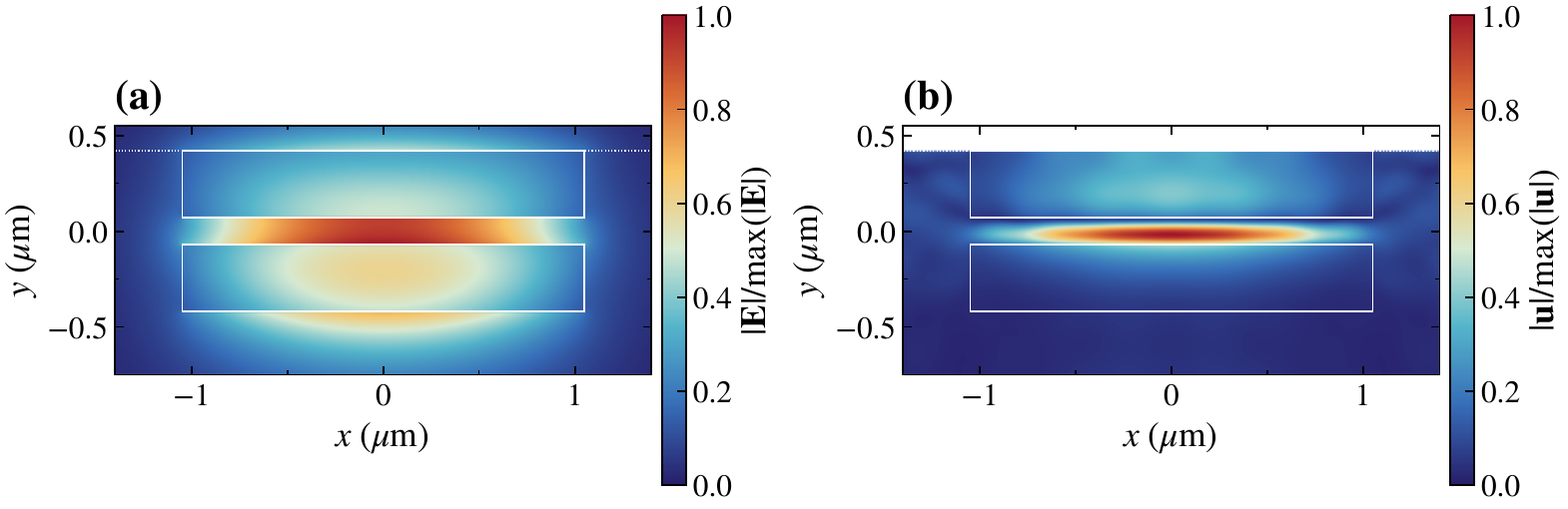}
\caption{(a) Confinement optical mode with $n_{\text{eff}} = 1.659$ and $\lambda = 1550$ nm. (b) Confinement acoustic mode with $Q_m \approx 400$, $\Omega_m/(2\pi) \approx 18.13$~GHz. The waveguide parameters (Si$_3$N$_4$): $t = 350$ nm, $w = 2100$ nm. The gap between the two waveguides is $d = 140$ nm. The thicknesses of the cladding SiO$_2$ $t \approx 7$ $\mu$m.} 
\label{fig:2}
\end{figure}

The waveguide geometry was designed to support gap optical and acoustic modes and maximize overlap between them. The calculated optical mode is shown in Fig.~\ref{fig:2}a and the acoustic mode is shown in Fig.~\ref{fig:2}b.

\begin{figure}[htbp]
\centering\includegraphics[width=\linewidth]{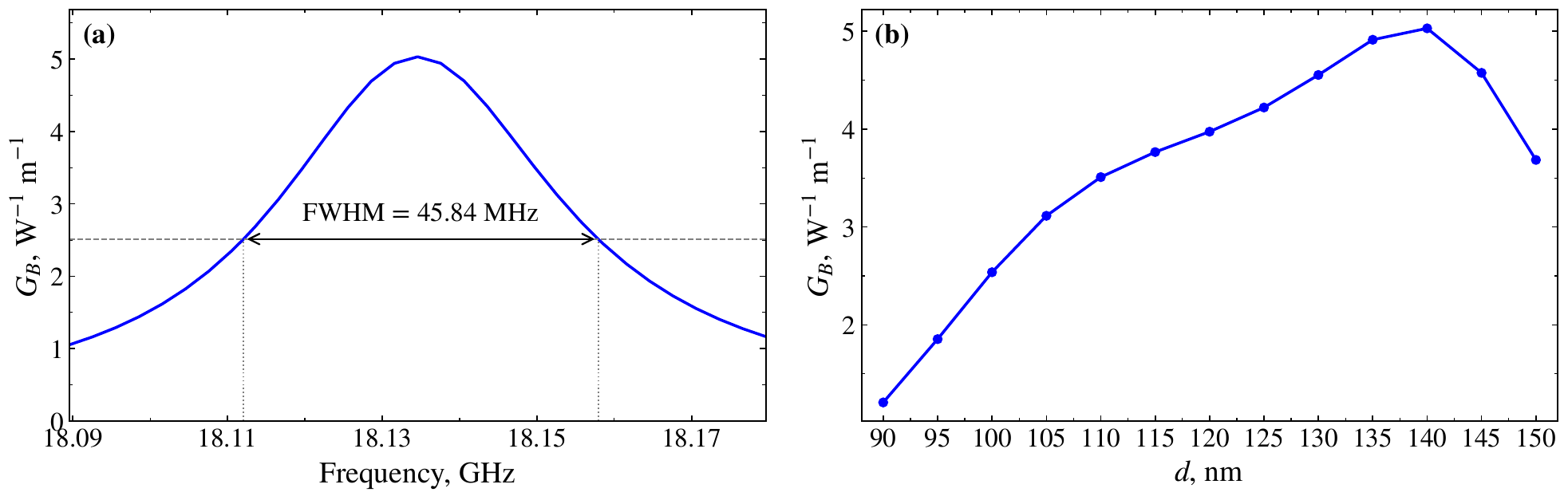}
\caption{(a) The Brillouin gain spectra calculated using Eqn. (\ref{eq:1}) near resonance acoustic mode Fig. \ref{fig:2}b. (b) The dependence of the peak Brillouin gain coefficient on the gap $d$. All parameters are the same as in Fig. \ref{fig:2}.} 
\label{fig:3}
\end{figure}

We find that at a given width and thickness of ${\rm{Si_3 N_4}}$ layers, there is the optimal gap $d$ corresponding to maximal Brillouin gain coefficient.  In the considered case $w=2100{~\rm{nm}}$, $t=350{~\rm{nm}}$, this optimal value $d = 140$ nm. The corresponding resonance frequency of the acoustic mode is around $\Omega_m/(2\pi) = 18.13$~GHz. The mechanical quality factor is $Q_m \approx 400$ (Fig. \ref{fig:3}a).

\begin{table*}[htbp]
\centering
\begin{tabular}{|c|c|c|c|c|}
\hline
        Work & Brillouin Gain, $G_B$ W$^{-1}$m$^{-1}$ & Frequency, GHz & Gain linewidth, MHz \\ \hline
        \cite{gyger2020observation} & $0.3$ & $25$ & $390$ \\ \hline
        \cite{botter2022guided} (Suppl. Mat.) & $1.2$ & $14$ & $35$ \\ \hline
        \cite{gundavarapu2019} & $0.1$ & $10.93$ & $153$ \\ \hline
        This work & $5$ & $18.13$ & 46 \\ \hline
\end{tabular}
\caption{Comparison of different works with integrated Si$_3$N$_4$ waveguides}
\label{tab:3}
\end{table*}

The comparison with results of other works is demonstrated in Table \ref{tab:3}. Here we can see that the subwavelength optical and acoustic modes theoretically give the highest Brillouin gain for such a chip-scale silicon nitride structure. 

The optimal gap width $d$ arises from the different confinement mechanisms of the optical and acoustic modes. The optical gap mode is formed by the interference of two TM-polarized modes supported by the Si$_3$N$_4$ waveguides. When the gap $d$ decreases, optical modes overlap more strongly, increasing the optical field intensity in the SiO$_2$ gap~\cite{almeida2004guiding}. In the opposite case, when the gap $d$ increases, the coupling between the waveguides weakens, leading to a field minimum near the center of the gap. This reduces the overlap with the acoustic mode, which is localized inside the gap of the SiO$_2$.

The acoustic mode is confined by reflections at the Si$_3$N$_4$ material. Decreasing the gap width $d$ initially improves acoustic confinement. However, when the gap becomes less than the transverse acoustic wavelength, $d<\lambda_{m, \perp}/2$, the acoustic mode can no longer be efficiently confined in the SiO$_2$ region, and its confinement weakens. The competition between these optical and acoustic confinement mechanisms results in an optimal gap width $d$ (Fig. \ref{fig:3}b).

\section{Conclusion}
In conclusion, we have demonstrated numerically the subwavelength confinement of optical and acoustic modes in the planar waveguide structure of two layers separated by the gap. The proposed geometry supports a hybridized TM optical mode and a strongly confined acoustic mode in the nanoscale gap. We have shown that the different confinement mechanisms of these modes lead to an optimal waveguide geometry that maximizes their spatial overlap. Using an example of the structure of two Si$_3$N$_4$ layers with a SiO$_2$ gap, we demonstrate that for the gap width of approximately $140$ nm and a layer's width of $2.1$ $\mu$m, the calculated Brillouin gain reaches $5$ W$^{-1}$m$^{-1}$ at an acoustic frequency of $18.13$ GHz.
The proposed structure enhances the opto-acoustic interaction without suspending the waveguide or introducing complex phononic structures, which opens a way for developing compact, low-power Brillouin-based devices for integrated photonics.

To enhance the Brillouin gain $G_B$ in Si$_3$N$_4$ integrated structures, previous studies have used single layer thin planar waveguides with thicknesses $40$~nm \cite{gundavarapu2019} that surrounded SiO$_2$ substrate. The small thickness of the Si$_3$N$_4$ layer leads to the deconfinement of the optical mode in the layer and ensures strong interaction with the acoustic modes in SiO$_2$. Furthermore, in \cite{botter2022guided} two thin Si$_3$N$_4$ waveguides are used to obtain a guided acoustic mode. For such a system to obtain hybrid TE mode, the thickness of the two Si$_3$N$_4$ planar waveguides has to be less than for the similar hybrid TM mode considered in Fig. \ref{fig:2}b. In \cite{botter2022guided} it is obtained that overlap between modes can be increased multiple orders of magnitude due to both confinement of the hybrid TE optical and acoustic modes. These structures can provide Brillouin gain $G_B$ from $0.3$ to $1.2$ W$^-1$ m$^-1$ (Table \ref{tab:3}). In our structure, a hybrid TM mode concentrates the optical field in a nanoscale SiO$_2$ gap, where it overlaps with the acoustic mode. This mechanism uses optical confinement in the gap for waveguide thicknesses of $350$~nm and above. This is an alternative approach to achieve strong opto-acoustic overlap without requiring an ultrathin waveguide, which can provide high Brillouin gain $G_B \approx 5$ W$^{-1}$ m$^{-1}$ (Table \ref{tab:3}).

\section*{Acknowledgments}
A.R.M., E.S.A. and A.A.Z. thank the foundation for the advancement of theoretical physics and mathematics “Basis”. The authors thank A.A. Elistratov for helpful discussion.

\nocite{*}

\bibliography{apssamp}

@PREAMBLE{
 "\providecommand{\noopsort}[1]{}" 
 # "\providecommand{\singleletter}[1]{#1}%" 
}

@article{mishra2026ultra,
  title={Ultra-low loss piezo-optomechanical low-confinement silicon nitride platform for visible wavelength quantum photonic circuits},
  author={Mishra, Mayank and Choi, Gwangho and He, Wenhua and Talcott, Gina M and Kearney, Katherine and Gehl, Michael and Leenheer, Andrew and Dominguez, Daniel and Otterstrom, Nils T and Eichenfield, Matt},
  journal={Nature Commun.},
  volume={},
  number={},
  pages={},
  year={2026}
}

@article{gyger2020observation,
  title={Observation of stimulated Brillouin scattering in silicon nitride integrated waveguides},
  author={Gyger, Flavien and Liu, Junqiu and Yang, Fan and He, Jijun and Raja, Arslan S and Wang, Rui Ning and Bhave, Sunil A and Kippenberg, Tobias J and Th{\'e}venaz, Luc},
  journal={Phys. Rev. Lett.},
  volume={124},
  number={1},
  pages={013902},
  year={2020},
  publisher={APS}
}

@article{wiederhecker2019brillouin,
  title={Brillouin optomechanics in nanophotonic structures},
  author={Wiederhecker, Gustavo S and Dainese, Paulo and Mayer Alegre, Thiago P},
  journal={APL Photonics},
  volume={4},
  number={7},
  pages={071101},
  year={2019},
  publisher={AIP Publishing}
}

@article{merklein2022100,
  title={100 years of Brillouin scattering: Historical and future perspectives},
  author={Merklein, Moritz and Kabakova, Irina V and Zarifi, Atiyeh and Eggleton, Benjamin J},
  journal={Appl. Phys. Rev.},
  volume={9},
  number={4},
  pages={041306},
  year={2022},
  publisher={AIP Publishing}
}

@article{eggleton2019brillouin,
  title={Brillouin integrated photonics},
  author={Eggleton, Benjamin J and Poulton, Christopher G and Rakich, Peter T and Steel, Michael J and Bahl, Gaurav},
  journal={Nature Photonics},
  volume={13},
  number={10},
  pages={664--677},
  year={2019},
  publisher={Nature Publishing Group UK, London}
}

@article{ren2025stimulated,
  title={Stimulated Brillouin scattering in micro/nanophotonic waveguides and resonators},
  author={Ren, Linhao and Wang, Wenyu and Xu, Kang and Zhu, Liying and Wang, Jun and Shi, Lei and Zhang, Xinliang},
  journal={Nanophotonics},
  volume={14},
  number={10},
  pages={1509--1535},
  year={2025},
  publisher={De Gruyter}
}

@article{rodrigues2025cross,
  title={Cross-polarized stimulated Brillouin scattering in lithium niobate waveguides},
  author={Rodrigues, Caique C and Schilder, Nick J and Zurita, Roberto O and Magalh{\~a}es, Let{\'\i}cia S and Shams-Ansari, Amirhassan and dos Santos, Felipe JL and Paiano, Ot{\'a}vio M and Alegre, Thiago PM and Lon{\v{c}}ar, Marko and Wiederhecker, Gustavo S},
  journal={Phys. Rev. Lett.},
  volume={134},
  number={11},
  pages={113601},
  year={2025},
  publisher={APS}
}

@article{rodrigues2023stimulated,
  title={Stimulated Brillouin scattering by surface acoustic waves in lithium niobate waveguides},
  author={Rodrigues, Caique C and Zurita, Roberto O and Alegre, Thiago PM and Wiederhecker, Gustavo S},
  journal={J.e Opt. Soc. Am. B},
  volume={40},
  number={5},
  pages={D56--D63},
  year={2023},
  publisher={Optica Publishing Group}
}

@article{van2016unifying,
  title={Unifying Brillouin scattering and cavity optomechanics},
  author={Van Laer, Rapha{\"e}l and Baets, Roel and Van Thourhout, Dries},
  journal={Phys. Rev. A},
  volume={93},
  number={5},
  pages={053828},
  year={2016},
  publisher={APS}
}

@article{GonzalezAndrade2024,
  author  = {Gonzalez-Andrade, David and Nuno Ruano, Paula and Zhang, Jianhao and Cassan, Eric and Marris-Morini, Delphine and Vivien, Laurent and Lanzillotti-Kimura, Norberto Daniel and Alonso-Ramos, Carlos},
  title   = {Enhancing stimulated {B}rillouin scattering in suspended silicon waveguides through subwavelength nanostructuration [Invited]},
  journal = {Opt. Mater. Express},
  volume  = {14},
  pages   = {2562},
  year    = {2024}
}

@article{NunoRuano2025,
  author  = {Nuno Ruano, Paula and Zhang, Jianhao and Gonzalez-Andrade, David and Melati, Daniele and Cassan, Eric and Cheben, Pavel and Vivien, Laurent and Lanzillotti-Kimura, Norberto Daniel and Alonso-Ramos, Carlos},
  title   = {Stimulated Forward {B}rillouin Scattering in Subwavelength Grating Silicon Membranes},
  journal = {Laser Photonics Rev.},
  year    = {2025},
  note    = {Early View}
}

@article{VanLaer2015,
  author  = {Van Laer, Rapha{\"e}l and Kuyken, Bart and Van Thourhout, Dries and Baets, Roel},
  title   = {Interaction between light and highly confined hypersound in a silicon photonic nanowire},
  journal = {Nat. Photonics},
  volume  = {9},
  number  = {3},
  pages   = {199--203},
  year    = {2015}
}

@article{zurita2021,
  title={Designing of strongly confined short-wave Brillouin phonons in silicon waveguide periodic lattices},
  author={Zurita, Roberto O and Wiederhecker, Gustavo S and Mayer Alegre, Thiago P},
  journal={Opt. Express},
  volume={29},
  number={2},
  pages={1736--1748},
  year={2021},
  publisher={Optical Society of America}
}

@article{Dainese2006,
  author  = {Dainese, P. and Russell, P. St. J. and Joly, N. and Knight, J. C. and Wiederhecker, G. S. and Laude, V.},
  title   = {Stimulated {B}rillouin scattering from multi-{GHz}-guided acoustic phonons in nanostructured photonic crystal fibres},
  journal = {Nat. Phys.},
  volume  = {2},
  number  = {6},
  pages   = {388--392},
  year    = {2006}
}

@article{gundavarapu2019,
  title={Sub-hertz fundamental linewidth photonic integrated Brillouin laser},
  author={Gundavarapu, Sarat and Brodnik, Grant M and Puckett, Matthew and Huffman, Taran and Bose, Debapam and Behunin, Ryan and Wu, Jianfeng and Qiu, Tiequn and Pinho, C{\'a}tia and Chauhan, Nitesh and others},
  journal={Nat. Photonics},
  volume={13},
  number={1},
  pages={60--67},
  year={2019},
  publisher={Nature Publishing Group UK London}
}

@article{botter2022guided,
  title={Guided-acoustic stimulated Brillouin scattering in silicon nitride photonic circuits},
  author={Botter, Roel and Ye, Kaixuan and Klaver, Yvan and Suryadharma, Radius and Daulay, Okky and Liu, Gaojian and van den Hoogen, Jasper and Kanger, Lou and van der Slot, Peter and Klein, Edwin and others},
  journal={Sci. Adv.},
  volume={8},
  number={40},
  pages={eabq2196},
  year={2022},
  publisher={American Association for the Advancement of Science}
}

@article{li2012characterization,
  title={Characterization of a high coherence, Brillouin microcavity laser on silicon},
  author={Li, Jiang and Lee, Hansuek and Chen, Tong and Vahala, Kerry J},
  journal={Opt. Express},
  volume={20},
  number={18},
  pages={20170--20180},
  year={2012},
  publisher={Optical Society of America}
}

@article{panyaev2024sub,
  title={Sub-100-Hz DFB laser injection-locked to PM fiber ring cavity},
  author={Panyaev, Ivan S and Itrin, Pavel A and Korobko, Dmitry A and Fotiadi, Andrei A},
  journal={J. Light. Technol.},
  volume={42},
  number={8},
  pages={2928--2937},
  year={2024},
  publisher={IEEE}
}

@article{chen2023optomechanical,
  title={Optomechanical ring resonator for efficient microwave-optical frequency conversion},
  author={Chen, I-Tung and Li, Bingzhao and Lee, Seokhyeong and Chakravarthi, Srivatsa and Fu, Kai-Mei and Li, Mo},
  journal={Nature Commun.},
  volume={14},
  number={1},
  pages={7594},
  year={2023},
  publisher={Nature Publishing Group UK, London}
}

@article{qiu2013stimulated,
  title={Stimulated Brillouin scattering in nanoscale silicon step-index waveguides: a general framework of selection rules and calculating SBS gain},
  author={Qiu, Wenjun and Rakich, Peter T and Shin, Heedeuk and Dong, Hui and Solja{\v{c}}i{\'c}, Marin and Wang, Zheng},
  journal={Opt. Express},
  volume={21},
  number={25},
  pages={31402--31419},
  year={2013},
  publisher={Optical Society of America}
}

@article{mukhamedyanov2026using,
  author={Mukhamedyanov, Artem R. and Zyablovsky, Alexander A. and Andrianov, Evgeny S.},
  journal={IEEE J. Sel. Topics Quantum Electron.}, 
  title={Using the Control/Driven Injection Locking Scheme in a Brillouin Optomechanical System for the Design of All-Optical Nonlinear Photonic Elements}, 
  year={2026},
  volume={32},
  number={5: Self-Injection Locked Lasers and Assoc. Sys.},
  pages={2200108-2200108},
  doi={10.1109/JSTQE.2026.3700852}}

@article{mukhamedyanov2026all,
title = {All-optical Brillouin random number generator},
journal = {Phys. Lett. A},
volume = {581},
pages = {131540},
year = {2026},
issn = {0375-9601},
doi = {https://doi.org/10.1016/j.physleta.2026.131540},
url = {https://www.sciencedirect.com/science/article/pii/S0375960126002161},
author = {Artem Mukhamedyanov and Evgeny Andrianov and Alexander Zyablovsky},
}

@article{almeida2004guiding,
  title={Guiding and confining light in void nanostructure},
  author={Almeida, Vilson R and Xu, Qianfan and Barrios, Carlos A and Lipson, Michal},
  journal={Opt. Lett.},
  volume={29},
  number={11},
  pages={1209--1211},
  year={2004},
  publisher={Optical Society of America}
}

@article{wolff2015,
  title={Stimulated Brillouin scattering in integrated photonic waveguides: Forces, scattering mechanisms, and coupled-mode analysis},
  author={Wolff, Christian and Steel, Michael J and Eggleton, Benjamin J and Poulton, Christopher G},
  journal={Physical Review A},
  volume={92},
  number={1},
  pages={013836},
  year={2015},
  publisher={APS}
}

@article{jin2021hertz,
  title={Hertz-linewidth semiconductor lasers using CMOS-ready ultra-high-Q microresonators},
  author={Jin, Warren and Yang, Qi-Fan and Chang, Lin and Shen, Boqiang and Wang, Heming and Leal, Mark A and Wu, Lue and Gao, Maodong and Feshali, Avi and Paniccia, Mario and others},
  journal={Nature Photonics},
  volume={15},
  number={5},
  pages={346--353},
  year={2021},
  publisher={Nature Publishing Group UK, London}
}

@article{ye2025integrated,
  title={Integrated Brillouin photonics in thin-film lithium niobate},
  author={Ye, Kaixuan and Feng, Hanke and Te Morsche, Randy and Wei, Chuangchuang and Klaver, Yvan and Mishra, Akhileshwar and Zheng, Zheng and Keloth, Akshay and Tar{\i}k I{\c{s}}{\i}k, Ahmet and Chen, Zhaoxi and others},
  journal={Science advances},
  volume={11},
  number={18},
  pages={eadv4022},
  year={2025},
  publisher={American Association for the Advancement of Science}
}

@article{godet2022micronewton,
  title={Micronewton nanofiber force sensor using Brillouin scattering},
  author={Godet, Adrien and Chretien, Jacques and Huy, Kien Phan and Beugnot, Jean-Charles},
  journal={Optics Express},
  volume={30},
  number={2},
  pages={815--824},
  year={2022},
  publisher={Optical Society of America}
}

@article{huang2017brillouin,
  title={Brillouin scattering from hybrid acoustic wave in a microscaled fiber for gas pressure sensing},
  author={Huang, Jinding and Zhong, Xiaoxuan and Liang, Hao and Cheng, Linghao and Li, Jie and Guan, Bai-Ou},
  journal={IEEE Photonics Journal},
  volume={9},
  number={2},
  pages={1--6},
  year={2017},
  publisher={IEEE}
}

@article{zhang2022refractive,
  title={Refractive index sensitivity of Brillouin acoustic modes in single-mode subwavelength-diameter fibers},
  author={Zhang, Yongsheng and Lu, Haifei and Deng, Shuo and Wen, Xiaoyan and Li, Min},
  journal={Applied Optics},
  volume={61},
  number={17},
  pages={5055--5061},
  year={2022},
  publisher={Optica Publishing Group}
}

@article{dong2015brillouin,
  title={Brillouin-scattering-induced transparency and non-reciprocal light storage},
  author={Dong, Chun-Hua and Shen, Zhen and Zou, Chang-Ling and Zhang, Yan-Lei and Fu, Wei and Guo, Guang-Can},
  journal={Nature communications},
  volume={6},
  number={1},
  pages={6193},
  year={2015},
  publisher={Nature Publishing Group UK, London}
}

@article{sohn2021electrically,
  title={Electrically driven optical isolation through phonon-mediated photonic Autler--Townes splitting},
  author={Sohn, Donggyu B and {\"O}rsel, O{\u{g}}ulcan E and Bahl, Gaurav},
  journal={Nature Photonics},
  volume={15},
  number={11},
  pages={822--827},
  year={2021},
  publisher={Nature Publishing Group UK, London}
}

@article{yang2024proposal,
  title={Proposal for Brillouin microwave-to-optical conversion on a chip},
  author={Yang, Yuan-Hao and Wang, Jia-Qi and Xu, Xin-Biao and Li, Ming and Zhang, Yan-Lei and Pan, Xiaoxuan and Xiao, Lintao and Wang, Weiting and Guo, Guang-can and Sun, Luyan and others},
  journal={Optical Materials Express},
  volume={14},
  number={10},
  pages={2400--2412},
  year={2024},
  publisher={Optica Publishing Group}
}

@article{slinkov2025all,
  title={All-optical nonlinear activation function based on stimulated Brillouin scattering},
  author={Slinkov, Grigorii and Becker, Steven and Englund, Dirk and Stiller, Birgit},
  journal={Nanophotonics},
  volume={14},
  number={16},
  pages={2711--2722},
  year={2025},
  publisher={De Gruyter}
}

@article{liu2025large,
  title={Large mode volume integrated Brillouin lasers for scalable ultra-low linewidth and high power},
  author={Liu, Kaikai and Nelson, Karl D and Behunin, Ryan O and Blumenthal, Daniel J},
  journal={Nature Communications},
  volume={16},
  number={1},
  pages={6419},
  year={2025},
  publisher={Nature Publishing Group UK, London}
}

@article{chauhan2021visible,
  title={Visible light photonic integrated Brillouin laser},
  author={Chauhan, Nitesh and Isichenko, Andrei and Liu, Kaikai and Wang, Jiawei and Zhao, Qiancheng and Behunin, Ryan O and Rakich, Peter T and Jayich, Andrew M and Fertig, C and Hoyt, CW and others},
  journal={Nature communications},
  volume={12},
  number={1},
  pages={4685},
  year={2021},
  publisher={Nature Publishing Group UK, London}
}

@article{poulton2013acoustic,
  title={Acoustic confinement and stimulated Brillouin scattering in integrated optical waveguides},
  author={Poulton, Christopher G and Pant, Ravi and Eggleton, Benjamin J},
  journal={Journal of the Optical Society of America B},
  volume={30},
  number={10},
  pages={2657--2664},
  year={2013},
  publisher={Optical Society of America}
}

@article{kittlaus2016large,
  title={Large Brillouin amplification in silicon},
  author={Kittlaus, Eric A and Shin, Heedeuk and Rakich, Peter T},
  journal={Nature Photonics},
  volume={10},
  number={7},
  pages={463--467},
  year={2016},
  publisher={Nature Publishing Group}
}

@article{rakich2012giant,
  title={Giant enhancement of stimulated Brillouin scattering in the subwavelength limit},
  author={Rakich, Peter T and Reinke, Charles and Camacho, Ryan and Davids, Paul and Wang, Zheng},
  journal={Physical Review X},
  volume={2},
  number={1},
  pages={011008},
  year={2012},
  publisher={APS}
}

@article{yu2025chip,
  title={On-Chip Brillouin Amplifier in Suspended Lithium Niobate Nanowaveguides},
  author={Yu, Simin and Zhou, Ruixin and Yang, Guangcanlan and Zhang, Qiang and Zhu, Huizong and Yang, Yuanhao and Xu, Xin-Biao and Chen, Baile and Zou, Chang-Ling and Lu, Juanjuan},
  journal={Laser \& Photonics Reviews},
  volume={19},
  number={16},
  pages={2500027},
  year={2025},
  publisher={Wiley Online Library}
}

@article{mayor2025high,
  title={High photon-phonon pair generation rate in a two-dimensional optomechanical crystal},
  author={Mayor, Felix M and Malik, Sultan and Primo, Andr{\'e} G and Gyger, Samuel and Jiang, Wentao and Alegre, Thiago PM and Safavi-Naeini, Amir H},
  journal={Nature communications},
  volume={16},
  number={1},
  pages={2576},
  year={2025},
  publisher={Nature Publishing Group UK, London}
}

@article{eichenfield2009optomechanical,
  title={Optomechanical crystals},
  author={Eichenfield, Matt and Chan, Jasper and Camacho, Ryan M and Vahala, Kerry J and Painter, Oskar},
  journal={nature},
  volume={462},
  number={7269},
  pages={78--82},
  year={2009},
  publisher={Nature Publishing Group UK, London}
}

@article{jiang2019lithium,
  title={Lithium niobate piezo-optomechanical crystals},
  author={Jiang, Wentao and Patel, Rishi N and Mayor, Felix M and McKenna, Timothy P and Arrangoiz-Arriola, Patricio and Sarabalis, Christopher J and Witmer, Jeremy D and Van Laer, Rapha{\"e}l and Safavi-Naeini, Amir H},
  journal={Optica},
  volume={6},
  number={7},
  pages={845--853},
  year={2019},
  publisher={Optical Society of America}
}

@article{cao2018surface,
  title={Surface-enhanced Raman scattering using nanoporous gold on suspended silicon nitride waveguides},
  author={Cao, Qipu and Feng, Jijun and Lu, Hongliang and Zhang, Hui and Zhang, Fuling and Zeng, Heping},
  journal={Optics Express},
  volume={26},
  number={19},
  pages={24614--24620},
  year={2018},
  publisher={Optical Society of America}
}

@article{mukhamedyanov2024hard,
  title={Hard excitation mode of a system with optomechanical instability},
  author={Mukhamedyanov, Artem and Zyablovsky, Alexander A and Andrianov, Evgeny S},
  journal={Optics Letters},
  volume={49},
  number={4},
  pages={782--785},
  year={2024},
  publisher={Optica Publishing Group}
}

@article{mukhamedyanov2025control,
  title={Control of the threshold of the hard excitation mode of an optomechanical system by a low-intensity seed wave},
  author={Mukhamedyanov, Artem R and Andrianov, Evgeny S and Zyablovsky, Alexander A},
  journal={Optics Letters},
  volume={50},
  number={19},
  pages={6121--6124},
  year={2025},
  publisher={Optica Publishing Group}
}

@article{mukhamedyanov2023subthreshold,
  title={Subthreshold phonon generation in an optomechanical system with an exceptional point},
  author={Mukhamedyanov, Artem and Zyablovsky, Alexander A and Andrianov, Evgeny S},
  journal={Optics Letters},
  volume={48},
  number={7},
  pages={1822--1825},
  year={2023},
  publisher={Optica Publishing Group}
}

\end{document}